\documentclass[conference]{IEEEtran}
\IEEEoverridecommandlockouts
\usepackage{cite}\usepackage{xcolor}
\usepackage{colortbl}
\usepackage{amsmath,amssymb,amsfonts}
\usepackage{algorithmic}
\usepackage{graphicx}
\usepackage{textcomp}
\usepackage{xcolor}
\usepackage[hidelinks]{hyperref}
\usepackage{graphicx}
\usepackage{tikz}
\def\BibTeX{{\rm B\kern-.05em{\sc i\kern-.025em b}\kern-.08em
    T\kern-.1667em\lower.7ex\hbox{E}\kern-.125emX}}

\newcommand{\Cross}{$\mathbin{\tikz [x=2ex,y=2ex,line width=.3ex, red] \draw (0,0) -- (1,1) (0,1) -- (1,0);}$}

\newcommand{\Checkmark}{$\color{green}\scalebox{1.5}{\checkmark}$}

\begin{document}

\title{Automated Microservice Pattern Instance Detection Using Infrastructure-as-Code Artifacts and Large Language Models\\
\thanks{The author thanks his supervisors Filipe F. Correia (FEUP, INESC TEC), Neil B. Harrison (Utah Valley University), and Ademar Aguiar (FEUP, \mbox{INESC TEC}) for their significant assistance, guidance, and availability. \\
This work is co-financed by Component 5 - Capitalization and Business Innovation, integrated into the Resilience Dimension of the Recovery and Resilience Plan within the scope of the Recovery and Resilience Mechanism (MRR) of the European Union (EU), framed in the Next Generation EU, for the period 2021 - 2026, within project HfPT, with reference 41.
}
}

\author{\IEEEauthorblockN{Carlos Eduardo Duarte}
\IEEEauthorblockA{\textit{Faculdade de Engenharia da Universidade do Porto} \\
\textit{Instituto de Engenharia de Sistemas e Computadores, Tecnologia e Ciência}\\
Porto, Portugal \\
ceduarte@fe.up.pt}
}

\maketitle

\begin{abstract}
Documenting software architecture is essential to preserve architecture knowledge, even though it is frequently costly. Architecture pattern instances, including microservice pattern instances, provide important structural software information. Practitioners should document this information to prevent knowledge vaporization. However, architecture patterns may not be detectable by analyzing source code artifacts, requiring the analysis of other types of artifacts. Moreover, many existing pattern detection instance approaches are complex to extend.

This article presents our ongoing PhD research, early experiments, and a prototype for a tool we call MicroPAD for automating the detection of microservice pattern instances. The prototype uses Large Language Models (LLMs) to analyze Infrastructure-as-Code (IaC) artifacts to aid detection, aiming to keep costs low and maximize the scope of detectable patterns. Early experiments ran the prototype thrice in 22 GitHub projects. We verified that 83\% of the patterns that the prototype identified were in the project. The costs of detecting the pattern instances were minimal. These results indicate that the approach is likely viable and, by lowering the entry barrier to automating pattern instance detection, could help democratize developer access to this category of architecture knowledge.

Finally, we present our overall research methodology, planned future work, and an overview of MicroPAD's potential industrial impact.
\end{abstract}

\begin{IEEEkeywords}
software architecture, architecture patterns, architecture documentation, pattern identification, knowledge retrieval, microservices, microservice patterns\end{IEEEkeywords}

%-------------------------------

\section{Introduction} \label{introduction}
Krutchen, Lago, and van Vliet state that ``the main value of a software company is its intellectual value"~\cite{hutchison_building_2006}. Unforeseen events may cause the loss of implicit software knowledge. It is, thus, important to preserve that knowledge so that companies can keep their competitive advantage.

Documenting software can achieve that. Documentation is one of the most valuable artifacts in a software project \cite{robillard_-demand_2017}, providing details on system usage and operation~\cite{aghajani_software_2019}, relevant for improving system comprehension and maintenance~\cite{chen_empirical_2009}.

The absence of documentation or its poor quality may cause significant issues during the software development life-cycle~\cite{chen_empirical_2009}. Despite this, most practitioners do not adequately maintain documentation~\cite{chen_empirical_2009, forward_relevance_2002, aghajani_software_2020, lethbridge_how_2003}, partly due to its high costs~\cite{robillard_-demand_2017} and complexity~\cite{aghajani_software_2019, lethbridge_how_2003}. When they do maintain it, it is often unreliable~\cite{forward_relevance_2002, lethbridge_how_2003}. Automated documentation techniques may help deal with these problems~\cite{martraire_living_2019, aghajani_software_2020}.

Documenting software pattern instances helps manage architecture knowledge by allowing practitioners to understand better the solution they are trying to implement~\cite{fowler_patterns_2003}. This knowledge is valuable for improving comprehension of the architecture decisions of a software system.

Research and practice show that patterns solve problems in software design. These include both design patterns and architecture patterns. Some architecture patterns fit within a microservice architecture, characterized by the arrangement of applications in a set of small, loosely coupled, and interoperable services. These are called microservice patterns~\cite{richardson_what_nodate}.

Design patterns, such as those from the Gamma et al. catalog~\cite{gamma_design_1995}, have been detected and documented in an automated manner. Many architecture patterns point towards using a set of components that interact in a given way. Design patterns are similar and usually prescribe implementing a set of classes. However, architecture patterns are more challenging to detect as they are much more abstract than design patterns. It may not be easy nor possible to recognize the presence of such architecture pattern components and their interactions using source code alone. For instance, there are many microservice patterns, each expressing themselves in their own way in software artifacts. Thus, traditional approaches may not be a good fit for detecting them.

Novel technologies such as Large Language Models (LLMs) may improve this process \cite{su_hotgpt_2023}. They have powerful pattern-matching capabilities and are context-aware (critical given the many possibilities of implementing architecture patterns). LLMs have been trained on massive amounts of data and may take less effort to implement and make pattern detection more accessible. Furthermore, Infrastructure-as-Code (IaC) artifacts provide information that may help detect microservice pattern instances. In this article, we distinguish source code from IaC. Thus, it is worth exploring their detection effectiveness.

Detecting microservice pattern instances in an automated manner using only IaC files as input and passing them to an LLM should significantly improve the architecture knowledge of software systems.

Our research aims to validate the following hypothesis:

\begin{quote}
    \textbf{H}: \textit{Detecting microservice pattern instances in an automated manner using both IaC artifacts and LLMs is an accurate, precise, and generalizable approach for effectively improving architecture knowledge.}
\end{quote}

The following research questions are guiding our work:

\begin{itemize}
    \item \textbf{RQ1.} \textit{Which IaC artifacts commonly found in software repositories contain relevant microservice pattern information?}
    \item \textbf{RQ2.} \textit{How accurately and precisely can microservice pattern instances be identified from software repositories?}
    \item \textbf{RQ3.} \textit{How much does the automated detection of microservice pattern instances improve practitioners' architecture knowledge?}
\end{itemize}

%--------------------------------------

\section{Related work} \label{related-work}

Researchers have studied improving architecture knowledge by producing comparative studies of knowledge management tools and analyzing the evolution of such tools over the years \cite{tang_comparative_2010, capilla_10_2016}. Because detecting and documenting architecture patterns improves architecture knowledge, these contributions are important to address RQ3 effectively.

While multiple tools for automating the detection of software pattern instances exist
~\cite{keller_pattern-based_1999, 
diamantopoulos_dp-core_2016,
gueheneuc_demima_2008, 
antoniol_object-oriented_2001,
pan_machine_2023}, these only identify instances from the original Gamma et al. catalog and do not include architecture or microservice patterns.

Alternatively, some tools focus on retrieving more general architecture knowledge~\cite{zhang_software_2023, granchelli_microart_2017, fonseca_x-trace_2007, cuadrado_case_2008}, powered by static and dynamic analysis techniques. These follow a data-driven approach using information extracted from multiple artifact types, such as code, documentation, metadata, and architecture models. However, none of these tools can explicitly identify microservice pattern instances~\cite{jansen_software_2005}. Moreover, only half of these tools are automated. We found one approach that can detect five microservice pattern instances based on specific metrics~\cite{daniel_towards_2023}. Nonetheless, the approach is not easily generalizable to other microservice pattern instances and can only detect a few patterns. 

Regarding the detection of microservices, multiple approaches for doing so exist~\cite{de_paoli_microservices_2017, dustdar_qualitative_2020}. However, these surveys do not focus on detecting microservice pattern instances. A survey on detecting microservice patterns, specifically API microservice patterns, showed that available tools need to be combined to detect a large set of patterns, that no tool detects patterns without human intervention, and that none use LLMs for detecting them~\cite{bakhtin_survey_2022}.

The use of LLMs for tackling software engineering problems has recently increased substantially~\cite{jahic_state_2024, steffen_large_2024, ampatzoglou_mapping_2024}. Nevertheless, to our knowledge, no approaches currently leverage LLMs to detect microservice pattern instances.

While research provides some insights on detecting microservice pattern instances (RQ2), we found a research gap regarding which IaC artifacts that are most common in code repositories contain relevant microservice pattern information (RQ1) and to what extent detecting these pattern instances improves practitioner architecture knowledge (RQ3). Thus, an approach for detecting a large pool of microservice pattern instances powered by LLMs and IaC artifacts may be a practical approach to improving practitioner architecture knowledge.

%---------------------------------------

\section{Proposed solution and methodology} \label{methodology}

LLMs may have issues when detecting microservice pattern instances. For large repositories, costs will undoubtedly increase tremendously (if using the best available models), so minimizing the amount of files sent to LLMs is necessary. Also, sending code artifacts to LLMs may pose a security and privacy risk when dealing with closed-source software because the code becomes part of the LLM's knowledge base. The amount of artifacts passed to the LLM should be surgically selected to reduce costs. Infrastrucure-as-Code (IaC) artifacts such as Terraform and Docker files allow defining and deploying infrastructure resources in an automated and consistent manner. These artifacts may be a cost-effective solution for detecting pattern instances. Additionally, these artifacts do not usually contain critical business logic, preventing jeopardizing their competitive advantage through repository leaks.

Considering this, we formalized the aforementioned research questions (RQs). Each RQ calls for different research methods. We based the research methods mentioned in this section on the ACM Empirical Standards~\cite{noauthor_acm_nodate} catalog.

For \textbf{RQ1}, we are performing \textit{Repository Mining} to analyze information related to identifying architecture pattern instances extracted from software repositories. In this context, \textit{information} is any insights an artifact can provide that help the LLM detect the pattern instances. We are looking for IaC artifacts contained in open-source GitHub repositories, which we manually inspect, prioritizing large repositories with high artifact type variety that implement a set of architecture patterns determined by the research group. This lookup enables us to infer which artifacts common in software repositories contain relevant information for detecting microservice patterns.

For \textbf{RQ2}, we are using \textit{Engineering Research} to develop a tool named \textit{MicroPAD}, which stands for Microservice Patterns Automated Detector, for detecting microservice architecture pattern instances from IaC artifacts. The results from RQ1 have a direct influence on our approach, impacting the information extraction strategy. We will validate the approach by an \textit{Experiment} where we will run the developed tool on repositories containing microservice architecture pattern instances that our approach can detect. Then, we will assess how close the identified pattern instances are to the ones present in the system. The research team will manually identify the pattern instances to compare the results of the approach to a single source of truth. We already finished developing a prototype for MicroPAD, and we give some initial results in Section \ref{preliminary-experiments}.

Finally, for \textbf{RQ3}, we will perform an \textit{Experiment} where we will ask a set of participants from both industry and academia about which pattern instances they know are present in the system (allowing participants to analyze system artifacts freely). Then, we will compare that with the results obtained from MicroPAD. After that, we will do a \textit{Qualitative Survey} with participants to collect feedback on how the results provided by MicroPAD changed their perception and knowledge of the system's architecture and how much better MicroPAD can detect patterns compared to practitioners.

%---------------------------------------------
\section{Preliminary work and experiments} \label{preliminary-experiments}

In order to verify the viability of our idea, we built a prototype for \mbox{MicroPAD} and experimented using it with a subset of patterns from Chris Richardson's catalog~\cite{richardson_what_nodate}. These experiments focus on understanding if the pattern instances that the prototype finds genuinely exist in the repository. We are not yet assessing whether the prototype found all pattern instances in the repository and contemplated in our catalog. Thus, at this point, pattern instances detected by the prototype do not necessarily mean other pattern instances are not present.

\subsection{Methodology}

We used OpenAI's GPT-4o mini model, which, while not considered the best LLM available at the time of publication, we found it to be the most cost-effective. We experimented with other models, such as GPT-4o (OpenAI's top-of-the-line model as of October 2024), but quickly found that the costs were much higher. Passing all repository files could increase costs exponentially and quickly hit rate limits. We also tested Gemini in its many variants but found the results lacking. Additionally, we attempted to use Small Language Models (SLMs), which allow for the local running of models. However, we found that they provided much worse and slower results and thus did not include them in the Experiment.

We selected the repositories used for testing by searching for the tag ``microservices" on GitHub. We chose the top 30 repositories with the most stars on the platform. We defined exclusion criteria, such as ignoring repositories that are not code projects (meaning they do not just contain pointers to resources) and hitting the LLM's rate limits. The exclusion criteria decreased the amount of analyzable repositories to 22.

We considered each repository's detected pattern instances in all three runs. Then, we manually analyzed whether the instances were present for each repository and each detected subset of pattern instances. Because LLM output is not deterministic, we ran the prototype on each repository thrice. The total costs accounted for 2 USD, which is affordable.

\subsection{Approach}
\begin{figure}[h]
    \centering
    \includegraphics[width=0.5\textwidth]{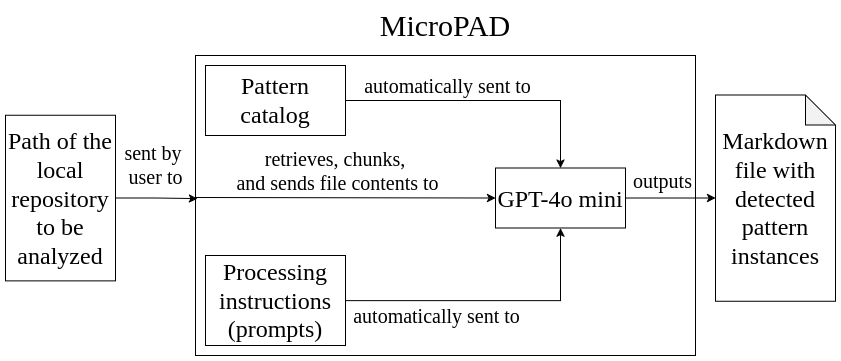}
    \caption{A brief overview of how the MicroPAD prototype works.}
    \label{fig:micropad-flow}
\end{figure}

MicroPAD's prototype was implemented using Python as a simple command-line tool and runs on repositories stored locally. The tool connects to ChatGPT's API and interacts with it by sending prompts, as shown in Fig.~\ref{fig:micropad-flow}. The user manually provides one primary input: the repository's path. The prototype parses all artifacts from that repository, which are then filtered. We consider the artifacts to be IaC artifacts if they match one of our pre-selected file extensions: .yml, .yaml, .dockercompose, .tf, .ps1, .pl, .properties, .json, .env, .sh, .kubeconfig, .helm, .tpl, .packer, .nomad, .vagrantfile, .ansible, .cloudformation, .template, .lock, .tfvars, .terraformignore, .tfplan, .env.example, .makefile, .cfg, .stack, .tfstate; or if they have specific names: Dockerfile or KubeFile.

To avoid LLM rate limits, we have split artifacts into chunks sent sequentially in separate prompts. This approach is not interactive: the only task of the prototype is to generate the output. Thus, the tool does not let the user send and receive messages to the LLM as if they were text chats.

The prototype also automatically sends a pattern catalog to the LLM as a Python file, with a set of strings containing information on each pattern and a set of instructions so that the LLM knows how to process the input to detect the pattern instances. The LLM uses a subset of Chris Richardson's catalog containing 24 out of the 52 presented patterns. The 24 patterns are: \textsc{Service instance per VM}; \textsc{Service instance per container}; \textsc{Sidecar}; \textsc{Service mesh}; \textsc{Service deployment platform}, \textsc{Single service per host}; \textsc{Serverless deployment}; \textsc{Client-side discovery}; \textsc{Self-registration}; \textsc{Service registry}; \textsc{Server-side service discovery}; \textsc{3rd party registration}; \textsc{API gateway/Backends for frontends}; \textsc{Circuit breaker}; \textsc{Remote procedure invocation}; \textsc{Messaging}; \textsc{Log deployments and changes}; \textsc{Log aggregation}; \textsc{Health check API}; \textsc{Application metrics}; \textsc{Audit logging}; \textsc{Shared database}; and \textsc{Database per service}.

Finally, the LLM outputs a Markdown file with the identified pattern instances, IaC code snippets supporting the detection, a justification on why the instances are present, and the LLM's detection certainty, ranging between High, Medium, and Low.

\subsection{Results}

For the sake of concision, we will only list pattern instances the prototype found with a high degree of certainty.

Table \ref{table:1} highlights which pattern instances were found to be present in IaC artifacts (signaled with a green checkmark) and which were not (signaled with a red cross). 

\begin{table*}[]
\centering
\caption{The MicroPAD prototype experiment results. We listed the names of the analyzed projects and the pattern instances detected in each of the three runs.}
\renewcommand{\arraystretch}{1.5}
\resizebox{.8\textwidth}{!}{%
\begin{tabular}{|l|p{5cm}|p{5cm}|p{5cm}|}
\hline
\textbf{Project} & \textbf{Run 1} & \textbf{Run 2} & \textbf{Run 3} \\ \hline

dubbo & 
 & 
\begin{tabular}[c]{@{}l@{}}- Service mesh \Cross \\ - Service instance per container \Cross \end{tabular} & 
\begin{tabular}[c]{@{}l@{}}- Service deployment platform \Checkmark \\ - Service registry \Checkmark\end{tabular} \\ \hline

go-zero & 
\begin{tabular}[c]{@{}l@{}}- Service instance per container \Checkmark \\ - Service deployment platform \Checkmark\end{tabular} & 
\begin{tabular}[c]{@{}l@{}}- Service instance per container \Checkmark \\ - Client-side discovery \Checkmark\end{tabular} & 
\begin{tabular}[c]{@{}l@{}}- Service instance per container \Checkmark \\ - Service deployment platform \Checkmark\end{tabular} \\ \hline

apollo & 
- Service instance per container \Checkmark & 
\begin{tabular}[c]{@{}l@{}}- Service instance per container \Checkmark \\ - Service deployment platform \Checkmark\end{tabular} & 
- Service instance per container \Checkmark \\ \hline

spring-cloud-alibaba & 
\begin{tabular}[c]{@{}l@{}}- Service Instance per Container \Checkmark \\ - API Gateway/Backends for Frontends \Checkmark \\ - Service registry \Checkmark \\ - Client-side discovery \Cross \\ - Database per service \Cross\end{tabular} & 
\begin{tabular}[c]{@{}l@{}}- Service deployment platform \Checkmark \\ - Service registry \Checkmark \\ - API Gateway/Backends for Frontends \Checkmark \\ - Service instance per container \Checkmark\end{tabular} & 
\begin{tabular}[c]{@{}l@{}}- Service instance per container \Checkmark \\ - Client-side discovery \Cross \\ - Service registry \Checkmark \\ - API Gateway/Backends for Frontends \Checkmark \\ - Database per service \Cross\end{tabular} \\ \hline

kit & 
\begin{tabular}[c]{@{}l@{}}- Service registry \Checkmark \\ - Client-side discovery \Checkmark \\ - Service Deployment Platform \Checkmark\end{tabular} & 
\begin{tabular}[c]{@{}l@{}}- Service registry \Checkmark \\ - Self-registration \Checkmark \\ - Client-side discovery \Checkmark\end{tabular} & 
\begin{tabular}[c]{@{}l@{}}- Service registry \Checkmark \\ - Self-registration \Checkmark \\ - Client-side discovery \Checkmark\end{tabular} \\ \hline

kratos & 
- Service registry \Checkmark & 
- Service registry \Checkmark & 
- Service instance per container \Checkmark \\ \hline

Sentinel & 
\begin{tabular}[c]{@{}l@{}}- Service instance per container \Checkmark \\ - Service deployment platform \Checkmark \end{tabular} & 
 & 
- Service instance per container \Checkmark \\ \hline

go-micro & 
- Service deployment platform \Checkmark & 
 & 
 \\ \hline

grpc-go & 
- Service instance per container \Checkmark & 
\begin{tabular}[c]{@{}l@{}}- Service instance per container \Checkmark \\ - Service deployment platform \Checkmark \\ - Client-side discovery \Checkmark \end{tabular} &
- API Gateway/Backends for Frontends \Checkmark \\ \hline

chi & 
\begin{tabular}[c]{@{}l@{}}- Service instance per container \Checkmark \\ - API Gateway/Backends for Frontends \Checkmark\end{tabular} & 
 & 
\begin{tabular}[c]{@{}l@{}}- Service instance per container \Checkmark \\ - API Gateway/Backends for Frontends \Checkmark\end{tabular} \\ \hline

zheng & 
- Service deployment platform \Checkmark & 
 & 
- Single Service per Host \Cross \\ \hline

quiankun & 
\begin{tabular}[c]{@{}l@{}}- Service instance per container \Cross \\ - API Gateway/Backends for Frontends \Cross \end{tabular} & 
- Service instance per container \Cross  & 
- Service deployment platform \Checkmark \\ \hline

jib & 
- Service instance per container \Checkmark & 
- Service instance per container \Checkmark & 
- Service instance per container \Checkmark \\ \hline

single-spa & 
- API Gateway/Backend for Frontends \Cross & 
- API Gateway/Backends for Frontends \Cross & 
\begin{tabular}[c]{@{}l@{}}- API Gateway/Backends for Frontends \Cross \\ - Service Deployment Platform \Checkmark\end{tabular} \\ \hline

piggymetrics & 
\begin{tabular}[c]{@{}l@{}}- Service instance per container \Checkmark \\ - Service deployment platform \Checkmark \\ - API Gateway/Backends for Frontends \Checkmark\end{tabular} & 
\begin{tabular}[c]{@{}l@{}}- Service deployment platform \Checkmark \\ - Service instance per container \Checkmark\end{tabular} & 
\begin{tabular}[c]{@{}l@{}}- Service instance per container \Checkmark \\ - Service registry \Checkmark \\ - API Gateway/Backends for Frontends \Checkmark \\ - Database per service \Checkmark\end{tabular} \\ \hline

kubeshark & 
\begin{tabular}[c]{@{}l@{}}- Service instance per container \Checkmark \\ - Service deployment platform \Checkmark \\ - Self-registration \Checkmark \\ - Service registry \Checkmark\end{tabular} & 
\begin{tabular}[c]{@{}l@{}}- Service instance per container \Checkmark \\ - Multiple services per host \Checkmark\\ - Service deployment platform \Checkmark\end{tabular} & 
\begin{tabular}[c]{@{}l@{}}- Service instance per container \Checkmark \\ - Service deployment platform \Checkmark\end{tabular} \\ \hline

Activiti & 
- Service deployment platform \Checkmark & 
- Service deployment platform \Checkmark & 
- Serverless deployment \Cross  \\ \hline

falcon & 
- Service deployment platform \Checkmark & 
\begin{tabular}[c]{@{}l@{}}- Service instance per container \Checkmark \\ - Health check API \Cross\end{tabular} & 
- Service instance per container \Checkmark \\ \hline

up & 
 & 
- Service deployment platform \Checkmark & 
- Service instance per container \Checkmark \\ \hline

karate & 
\begin{tabular}[c]{@{}l@{}}- Service instance per container \Checkmark \\ - Service deployment platform \Checkmark\end{tabular} & 
- Service instance per container \Checkmark & 
\begin{tabular}[c]{@{}l@{}}- Service instance per container \Checkmark \\ - Service deployment platform \Checkmark\end{tabular} \\ \hline

rpcx & 
\begin{tabular}[c]{@{}l@{}}- Service deployment platform \Checkmark \\ - Remote procedure invocation \Checkmark\end{tabular} & 
- Service deployment platform \Checkmark & 
\begin{tabular}[c]{@{}l@{}}- Service deployment platform \Checkmark \\ - Remote procedure invocation \Checkmark\end{tabular} \\ \hline

oatpp & 
 & 
 & 
 \\ \hline

\end{tabular}%
}
\label{table:1}
\end{table*}

The table cell is blank when the prototype found no patterns. The pattern instances in this table are those the tool considered to be present with a high degree of certainty. A cloud folder with all the pattern instances the MicroPAD prototype found with all degrees of certainty is available in the following link: \url{https://tinyurl.com/micropad-logs}. The pattern instances found for each repository in separate executions of the prototypes, with the same inputs, are indicated in columns \textit{Run 1}, \textit{Run 2}, and \textit{Run 3}.

We validated the correctness of the identified patterns by doing a manual analysis. The analysis involved studying the artifacts of the repositories listed in Table \ref{table:1} and determining if the identified patterns were, in fact, present. One of the researchers did this. We divided the number of all correctly detected pattern instances in all repositories and all three runs by the number of all detected pattern instances in all repositories and all three runs. We verified that 83\% of the patterns the prototype identified are in the project.

Early indicators are positive. Compared to existing approaches, the prototype is language agnostic \cite{diamantopoulos_dp-core_2016, pan_machine_2023, granchelli_microart_2017, de_paoli_microservices_2017, antoniol_object-oriented_2001}, making it more generalizable. The MicroPAD prototype does not require specific metrics or algorithms to detect each pattern, contrary to existing approaches \cite{gueheneuc_demima_2008, antoniol_object-oriented_2001, daniel_towards_2023}. Instead, it requires natural language descriptions of patterns, which should be more accessible. Our approach should also require less effort to implement and adapt than those using machine learning techniques, which may require labeling datasets and training models \cite{pan_machine_2023}. It is also affordable. The percentage of microservice pattern instances verified in repositories is promising and should continue to improve. Finally, these characteristics are important for industrial adoption and provide a unique and novel approach.

There are particular challenges. Factors such as the inconsistent output per run, the limitations of this study held by the premises we assumed at the beginning of this section, and a considerable portion of the patterns present in the catalog not being identified lead us to believe that we can still do more to answer RQ2 assertively. Alternatives provide excellent accuracy and precision values, something we cannot yet assert for our tool. There may be a tradeoff between generalizability and accuracy/precision, which could make it less appealing.

\section{Future research results and challenges} \label{future-work}
We want to further research RQ1. We will improve the LLM's input, as files analyzed by the LLM may pass our exclusion criteria but not contain much relevant information. We will examine repositories for pattern instances that were not detected. Moreover, we intend to study which IaC artifacts contain meaningful information for detecting microservice pattern instances.

For further research regarding RQ2, we will aim to minimize inconsistent instance detection. We will focus on understanding the characteristics of the detectable patterns by analyzing IaC artifacts. Moreover, we will also research which patterns MicroPAD finds most accurately and/or consistently. Furthermore, considering that the prototype only detected a subset of the pattern catalog used for the experiments with high certainty, we intend to research why that happened, even if they may be undetectable from IaC artifacts alone. We aim to validate the preliminary results further by using a larger pool of repositories and ensuring an independent verification of the detected pattern instances performed by more people. We also intend to detect a more extensive set of instances.

Regarding RQ3, we aim to improve the MicroPAD prototype's output from a Markdown file to a web application to make interaction more accessible. Finally, we want to research how much faster MicroPAD detects pattern instances.

We anticipate some challenges. We could obtain low-quality architecture information, which would compromise the correct detection of pattern instances. The incorrect detection could be further exacerbated by having limited access to high-quality repositories and our capacity to manually identify and label enough architecture pattern instances to be detected by our prototype. Moreover, the chosen approach (which depends on the answers to some of the research questions) may not be readily generalizable to more than just a subset of patterns, which could limit the tool's usefulness and assess if the produced documentation is helpful to practitioners.

%------------------------------

\section{Industrial Impact} \label{industrial-impact}

Researchers and practitioners train LLMs on enormous datasets. Thus, LLMs know many microservice pattern implementations and their context. Because MicroPAD has its pattern catalog, it prevents LLMs from going astray and enables them to use all of their knowledge. Consequently, MicroPAD may reduce the effort and complexity of detecting pattern instances while preserving architecture knowledge.

Adding patterns to MicroPAD's catalog is easy: adding the pattern description in a pattern catalog file using natural language is only necessary. There is also no need to manage and deploy an LLM. The simplicity of the process should help democratize access and usage of pattern instance detection tools, significantly lowering the technical and non-technical knowledge entry barrier.

We also designed our approach to be fast. Because LLMs run frequently on the cloud, they can process large amounts of data in seconds. Speed is critical for industrial adoption, especially when code bases are extensive. Additionally, MicroPAD removes the need to obtain and label datasets and train large learning models, which is often time-consuming. Furthermore, it is easy to start using MicroPAD quickly: providing an API key of the chosen LLM is necessary. Even though the highest-tier LLMs are not as affordable as possible, we obtained these results by spending minimal funds. As technology continues to evolve, we expect existing models to become cheaper and better models to become available, which would undoubtedly improve the precision and accuracy of our tool without significant code changes. Nevertheless, considering the obtained results and the cost of execution, MicroPAD provides good results. 

MicroPAD only uses IaC artifacts, which should not contain critical business logic. Only using these artifacts allows MicroPAD to detect pattern instances without jeopardizing the company's competitive advantage, which is often the software itself. LLMs may not always ensure the privacy of the input they receive, so this is especially relevant. Thus, our approach minimizes privacy issues while still leveraging the power of LLMs. However, if organizations still do not trust an LLM to process their code, SLMs may also be used. Even though their results are not as good as LLMs, as technology evolves, this may become increasingly viable, reducing privacy concerns.

%-------------------------------

\section{Conclusion} \label{conclusion}
We plan on continuing this research by tackling the points stated in Section \ref{future-work}. MicroPAD should give practitioners important architecture knowledge by detecting microservice pattern instances in an automated manner. MicroPAD could decrease the cost and effort of maintaining architecture knowledge. Our early results indicate that the tool may be accurate, practical, affordable, fast, and simple while requiring minimal resources and maximizing data privacy. Ultimately, MicroPAD should better position software companies to preserve their competitive advantage: that is, their intellectual capital.

\bibliographystyle{ieeetr}
\bibliography{icsa-refs}

\end{document}